# Isotope Effect on the Thermal Conductivity of Graphene

Hengji Zhang [1], Geunsik Lee [1], Alexandre F. Fonseca [2], Tammie L. Borders [3], Kyeongjae Cho [1,4,*]

1 Department of Physics, University of Texas at Dallas, Richardson, Texas 75080, USA

2 Departamento de Física, Instituto de Ciências Exatas (ICEx), Universidade Federal Fluminense, Volta Redonda, RJ, 27213-350, Brazil

3 Department of Chemistry, University of North Texas, Denton, Texas 76203

4 Department of Materisals Science and Engineering, University of Texas at Dallas, Richardson, Texas 75080, USA

*corresponding author: kjcho@utdallas.edu

## Abstract

The thermal conductivity (TC) of isolated graphene with different concentrations of isotopes ($C_{13}$) is studied with equilibrium molecular dynamics method at 300K. In the limit of pure $C_{12}$ or $C_{13}$ graphene, TC of graphene in zigzag and armchair directions are ~630 W/mK and ~1000W/mK, respectively. We find that the TC of graphene can be maximally reduced by ~80%, in both armchair and zigzag directions, when a random distribution of $C_{12}$ and $C_{13}$ is assumed at different doping concentrations. Therefore, our simulation results suggest an effective way to tune the TC of graphene without changing its atomic and electronic structure, thus yielding a promising application for nanoelectronics and thermoelectricity of graphene based nano-device.

## 1. Introduction:

Since it was fabricated in 2004 [1], graphene, a monolayer of sp2-bonded network of carbon atoms, has attracted much attention for its unique electronic properties [2]. Meanwhile, both recent theory and experiment studies [3, 4] have revealed that isolated graphene has shown an unusual high thermal transport capability, which is of great importance in thermal management of nanonelectronics. Different from conventional metallic materials, thermal energy carriers for graphene are mainly in the form of phonon vibrations [3-5], and phonon contribution to TC is approximately 50 times larger than electron contribution at room temperature [3], which suggests that electron thermal transport is negligible in our case. Furthermore, the electronic contribution to TC would be independent of isotope effect as $C_{13}$ is electronically identical to $C_{12}$. For these reasons, we will study only the phonon contribution to TC in this paper.

In addition to utilizing its high TC, another possible application of graphene has been investigated for thermoelectric energy conversion [6], where low TC but high electric conductivity for graphene is required for obtaining high thermoelectric efficiency. Such efficiency is expressed as $ZT = \frac{\sigma S^2}{\kappa} T$, where σ is the electric conductivity, S is the Seebeck coefficient, and $\kappa$ is the thermal conductivity. To achieve high ZT for graphene, a general scheme is to minimize $\kappa$ while keeping σ and S less changed. One practical method is to dope the graphene with the stable isotope $C_{13}$ since electronic structure of graphene is unchanged in this doping. In a pure crystal, one without defects or dislocations, phonon scattering in the presence of different isotopes has been strongly correlated with changes in thermal conductivity. Similar to the observation for isotope effects on TC of Ge, diamond, and boron nitride nanotubes [7, 8], it is interesting to check how effectively isotope doping method will reduce TC of Graphene.

Modeling of thermal transport can be achieved by using Bolzmann transport equation (BTE) [9, 10], or molecular dynamics (MD) simulations [11]. In BTE method, the single mode relaxation time (SMRT) approximation is a commonly used technique involving the assignment of the relaxation time to different phonon scattering mechanisms. The relaxation time can be either fitted to the experimental TC value [12] or determined from MD simulations [10]. In MD simulation approach, TC can be predicted from either nonequilibrium MD [13], where a temperature is applied across the simulation cell, or equilibrium MD [14], where the so-called Green-Kubo method is used to compute TC from heat current fluctuations. One advantage for MD method is that there is no assumption needed for phonon interactions. As long as the phonon dispersion and anharmonicity of the potential are accurate, MD method provides a robust way to accurately compute thermal transport.

In Section 2, we introduce Green-Kubo method and describe the simulation procedures to compute TC. In Section 3, we show the isotope effects on thermal transport of graphene, and discuss the reason for TC reduction and its possible application for thermoelectric application. Section 4 presents a summary and conclusions.

2. Green-Kubo MD simulation Method:

The Green-Kubo formula [14] derived from linear response theory can express thermal conductivity tensor in terms of equilibrium heat current-current autocorrelation in the form,

$$\kappa_{\alpha\beta} = \frac{1}{\Omega k_B T^2} \int_0^{\tau_m} \left\langle \vec{J_\alpha}(\tau) \bullet \vec{J_\beta}(0) \right\rangle d\tau \qquad (1)$$

Where $\Omega$ is the system volume defined as the area of graphene multiplied with van der Waals thickness (3.35 Angstrom), $k_B$ is the Boltzmann constant, T is the system temperature, and $\tau_m$ is the time required to be longer than the time for current-current correlations to decay to zero [11]. $J_{\alpha,\beta}$ is denoted as the heat flux in α or β direction, its expression is commonly defined as,

$$\vec{J} = \frac{d}{dt} \sum_i r_i(t) E_i(t) \qquad (2)$$

$$E_i = \frac{1}{2} m v_i^2 + \phi_i \qquad (3)$$

Where $r_i$, $v_i$ and $E_i$ are the position, velocity and site energy of atom i respectively. $\phi_i$ is the potential energy at site i. Then, we use Hardy's definition [15],

$$\vec{J} = \sum_i [E_i \vec{v_i} + \sum_{j \neq i} \vec{r_{ij}}(\vec{F_{ij}} \bullet \vec{v_i})], \qquad (4)$$

where $r_{ij}=r_i-r_j$, $F_{ij}$ is the force exerted on atom i by atom j. One merit of Hardy's definition is that it is independent of pair-wise or many-body potential formulas. Based on the above equations, heat current at each step is recorded on disk as a quantity defined by atom positions, velocities, and inter-atomic forces, which can be extracted from atomistic simulations. The last step is to compute TC tensor with the known heat current in equation (1).

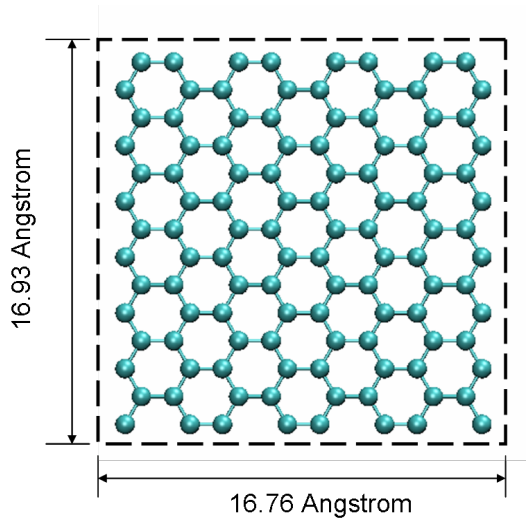

Fig.1 Structure of graphene unit cell with 112 carbon atoms.

In the equilibrium MD simulations, we used the second generation REBO carbon potential for its accuracy in describing bond strength and anharmonicity of carbon materials [16]. A unit cell of graphene, with a periodic boundary condition as shown in Fig.1, has been thermalized to 300K for 200 ps with Berendson thermostats. Afterwards, the heat current of graphene is recorded every 2fs in nine microcanonical ensemble simulations with uncorrelated initial conditions. Each microcanonical simulation needs to run up to 10 ns to obtain converged TC value. Compared with non-equilibrium molecular dynamics (NEMD), the Green-Kubo method is indeed less computationally efficient; however, it is free of troubles such as finite size and boundary effects, which are commonly unavoidable in NEMD.

3. Results and Discussion:

We computed the TC of mass defect free graphene as 630 W/mk and 1000 W/mk in armchair and zigzag direction, respectively. This is lower than the reported experimental data [4, 5]. The reason has to do with the discrepancies on phonon dispersion between experiment measurement and theoretical data predicted with original REBO potential [17]. Although the absolute TC of graphene is overall underestimated in our simulation, the relative difference of TC caused by isotope mass defect is still physically meaningful, and the normalized TC reduction seen in Fig. 4 properly reflects the isotope effect on TC of graphene.

One primary test done before studying isotope effects on thermal transport is the convergence test for graphene with different unit cell periodic boundary lengths. As shown in Fig.2, we have tested three cases with the boundary length of 1.6, 3.4 and 5.1 nm respectively, and the converged result suggests that the majority contribution from different phonon vibration modes are well included in our simulations. Then, to be efficient, we choose the 1.6 nm unit cell as the structure for the following simulations.

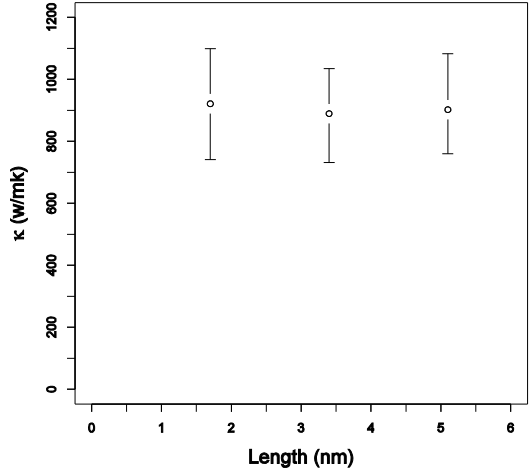

Fig. 2 Graphene TC convergence tests with different unit cell boundary lengths. κ is the average of TC in zigzag and armchair directions.

The first thing we learned from the simulations is that TC of pure $C_{12}$ or $C_{13}$ graphene in zigzag direction is ~58% greater than that in armchair direction. This can be conceptually explained from the acoustic phonon dispersion and Grüneisen parameter of graphene calculated with REBO potential as shown in Fig. 3. In the SMRT approximation, the BTE method can express TC as a summation of each phonon mode contribution, $\kappa = \sum_i \sum_q \kappa_i(\vec{q})$, and $\kappa_i = C_i(\vec{q}) v_i^2(\vec{q}) \tau^{ph}(\vec{q})$ where i and $\vec{q}$ represent phonon modes and wave momentum, $C_i(\vec{q})$ is the specific heat of the phonons, which is constant in the classical case. $v_i$ is the group velocity for mode i, $\tau^{ph}$ is the relaxation time and T is the temperature. From Fig.3 (a), we notice that both out of plane acoustic (ZA) and transverse acoustic (TA) modes in zigzag direction have apparently larger group velocity than that in armchair direction, longitudinal acoustic (LA) modes group velocity in both directions is quite similar. Then, from the Fig.3 (b), the Grüneisen parameter, which is plotted as a function of phonon mode i and momentum q, shows a smaller difference in zigzag and armchair directions for each vibration mode. As the Grüneisen parameters for LA, TA and ZA are similar, it is reasonable to assume that the relaxation times for phonon vibration in zigzag and armchair direction are the same. Based on the above analysis, we conclude that the difference in the TC between zigzag and armchair directions mainly comes from their different group velocities.

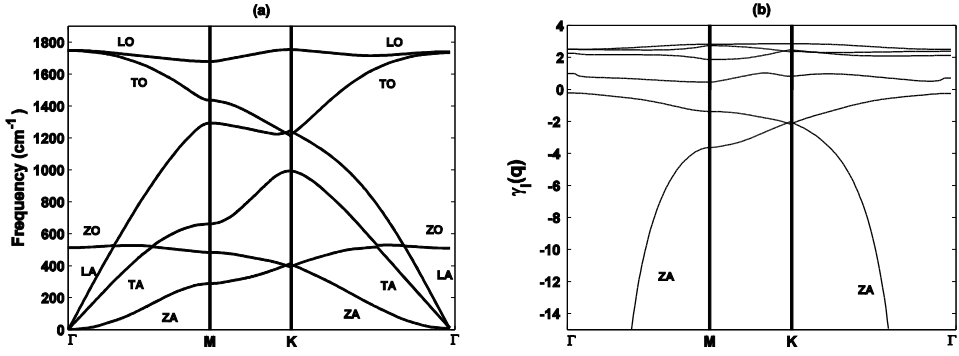

Fig. 3 a) Phonon dispersions for $C_{12}$ graphene and b) Grüneisen parameters of graphene as a function of q momentum for each vibration mode. ΓM and ΓK represent armchair and zigzag directions of graphene. Phonon dispersion for graphene is calculated with PHON code [18] for REBO carbon potential.

In the isotope effect study, we generated a wide range of graphene samples with different $C_{13}$ concentrations randomly distributed, since it is a more realistic possible configuration after the synthesis of graphene. Recently, Mingo *et al.* have proposed a possible method to generate isotope clusters in graphene, and theoretically demonstrated TC reduction using non-equilibrium Green's function method [19]. Fig.4 shows the normalized graphene TC values as a function of $C_{13}$ concentration in armchair and zigzag directions, and the maximum TC reduction for graphene can be made between 25% and 75% of the $C_{13}$ atomic concentrations.

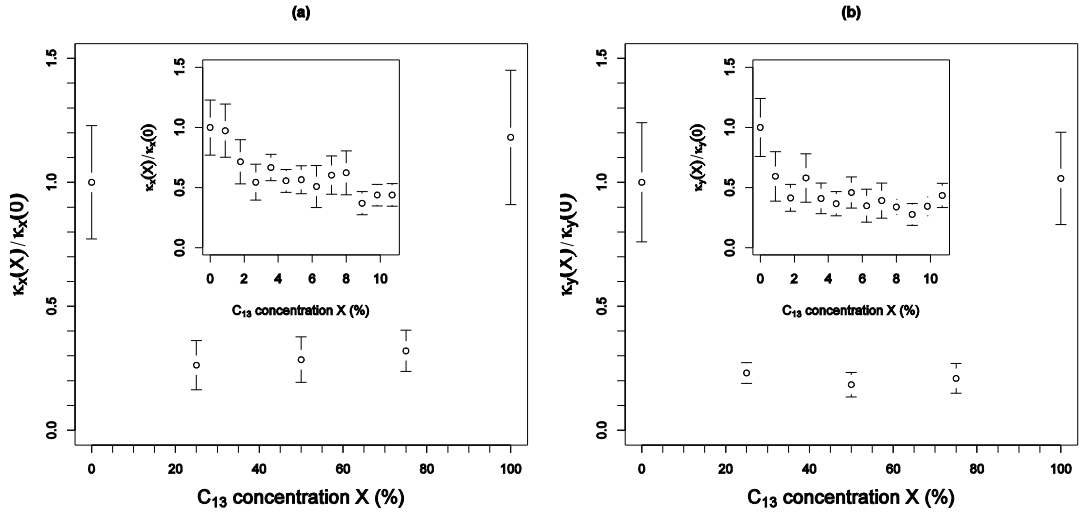

Fig. 4 Normalized Graphene TC as a function of $C_{13}$ concentration a) TC in armchair direction b) TC in zigzag direction. The inset shows TC as a function of $C_{13}$ for low concentration. The normalizing factors $\kappa_x$ and $\kappa_y$ are the values for pure graphene, 630 W/mK and 1000 W/mK, respectively.

The explanation of the almost "parabolic" shape of TC in Figure 4 can be found in an earlier important relation derived by Klemens [20]. As it is commonly accepted, the longer the phonon mean free path is, the larger TC would be. Klemens found that, for isotope scattering, the mean free path is proportional to $g^{-1}T^{-f}$, where $g = \sum_i C_i \left[ \dfrac{M_i^2 - (\sum C_i M_i)}{(\sum C_i M_i)^2} \right]^2$, and T is the temperature. $C_i$ and $M_i$ represent the concentration and the mass of the constituent isotope atoms, respectively. Thus, the mean free path directly has to do with the g factor, which is the mass variation of isotope atoms. In our simulation, g factor reaches its maximum for around 50% of $C_{13}$, so there we have the minimum phonon mean free path and the minimum TC. As $C_{13}$ atomic concentration approaches to zero or 100%, g becomes smaller, phonon free path is larger, and TC increases. It won't get to infinity because of the Umklapp and other phonon scattering mechanisms.

Among various methods that modulate the TC of graphene, the isotope doping method provides an efficient way to improve thermoelectric efficiency, since isotope atoms do not change the electronic structure of graphene, the electric conductivity and the Seebeck coefficient remain the same after the $C_{13}$ doping. As our simulations suggest, $\kappa$ in armchair and zigzag directions can be reduced by up to 80%, which means an improvement of the ZT by a factor of 5 since ZT is inversely proportional to TC. Recently, two important simulation studies were carried out on ZT of armchair [21] and zigzag [22] graphene nanoribbons (GNRs). In both works, the authors looked for defect and disordered structures that minimize TC while keep electric conductivity less changed, thus maximizing the ZT. Compared with this, the isotope doping method appears to be another attractive method that would promote ZT even further for GNRs when it is combined with those methods [21, 22]. We believe our results can motivate new experiments to check the efficiency of isotope doping method for thermoelectric application.

## 4. Conclusion:

In summary, we have studied isotope effect on TC of isolated graphene with equilibrium MD methods. Our simulation results suggest that TC of graphene can be effectively reduced by up to 80% in armchair and zigzag directions for isotope concentrations as low as 25%. The phenomenon that mass defect can reduce TC is explained with the relation between phonon mean free path and mass variation of the isotope mixtures [20]. Finally, our study shows that doping graphene with carbon isotope $C_{13}$ could be a practical way to reduce TC without changing its electric property, thus promoting thermoelectric coefficient.

## Acknowledgement:

The authors are grateful to support from Lockheed Martin. G.L. acknowledges support from SWAN, and A.F.F. acknowledges partial support from the Brazilian agency CNPq. We also appreciate discussions with Davide Donadio and Giulia Galli.